\documentclass[12pt]{article}
\usepackage[cp1251]{inputenc}  
\usepackage[english]{babel}
\usepackage{amsmath,amssymb,epsfig,graphicx,doc,latexsym,amsfonts,floatflt,bm,longtable}
\usepackage[bookmarksopen,colorlinks]{hyperref}
\usepackage[monochrome]{color}
 \hypersetup{colorlinks=true}

\begin{document}
\begin{center}
{\textbf{\LARGE{Temporal evolution of drift-type modes in plasma
flows with strong time-dependent velocity shear}}}

\bigskip
\bigskip
{\large{V.S. Mikhailenko$^{\dag,\ddag}$}}\footnote{Electronic mail:
\href{mailto:vmikhailenko@kipt.kharkov.ua}{vmikhailenko@kipt.kharkov.ua}},
{\large{V.V. Mikhailenko$^{\dag,\ddag}$, K.N.
Stepanov$^{\dag,\ddag}$}}
\\
\bigskip $^{\dag}${\textit{Kharkov National University, 61108 Kharkov,
Ukraine}}

$^{\ddag}${\textit{National Science Center ``Kharkov Institute of
Physics and Technology'',
\\
61108 Kharkov, Ukraine}}

\end{center}

\begin{abstract}
The linear and renormalized nonlinear analysis of the temporal
evolution of drift-type modes in plasma flows with strong
time-varying velocity shear is developed. Analysis is performed in
the time domain without spectral decomposition in time and admits
time variation of the flow velocity with time scales which may be
comparable with turbulent time scales.
\\ 52.35.Ra
\end{abstract}

\newpage

\section*{I. INTRODUCTION.}
A great variety of experimental results are now
available\cite{burell97-1499}, which confirm a suppression of
turbulence by $E\times B$ velocity shear as a key feature of  the
regimes of improved confinement\cite{Wagner} of plasma and formation
of transport barriers in tokamaks. The first remarkable success in
theoretical investigation of this fascinating and rich topic of
fusion research was achieved in Refs.\cite{Biglari,Shaing}, where
nonlinear enhanced decorrelation of fluctuations by shear flow had
been proposed as a universal and robust  mechanism for turbulence
quenching and transport reduction. It was compared quite favorably
with experimental data for numerous tokamaks and was confirmed by
gyrofluid codes\cite{Waltz1994, Waltz1995, Waltz1998}. This theory
was developed with the help of a simple "modal" approach, under an
implicit assumption that perturbations of fields, densities,
currents ets. are "normal eigenmodes" with certain frequencies and
wave numbers. However shear flow itself affects the structure of the
waves and their temporal evolution. The shear flow distorts wave
pattern, leads to its stretching and to temporal changing with time
the wave number of any separate wave\cite{Terry, Diamond}. Under
condition of strong flow shear, which is observed in edge layers of
tokamaks, where velocity shearing rate $dv_{0}/dr$ approaches or
even exceeds the drift wave frequencies, this theory has to be
strongly modified.

The experimentally relevant strong shear regime was studied
analytically in numerous papers applying the gyrokinetic approach
(see, for example,Refs.\cite{Brizard}, \cite{Hahm}; we mention also
Ref.\cite{Kawamura} as an example of a recent work). In our papers
(see Refs.\cite{Mikhailenko-2000}-\cite{Mikhailenko-2008}) the
investigations of linear evolution of drift waves in shear flows
applied Kelvin's method of shearing modes or a so-called non-modal
approach. This treatment is performed in terms of a physical time
variable without any prior spectral decomposition in time. Kelvin's
method appears to be very effective for the analytical
investigations of the temporal evolution of numerous low frequency
modes in plasma shear
flows\cite{Mikhailenko-2000}-\cite{Mikhailenko-2008}. Using one of
the simplest models available describing three--dimensional plasma
turbulence at the edge of a magnetic confinement device, the
Hasegawa--Wakatani equations\cite{Hasegava}, we find that under
condition of strong flow shear non-modal evolution of the drift--
type perturbations supersedes the modal development of linear
instabilities. Drift modes are suppressed before they can grow,
preventing the development of nonlinear processes and imposing its
own physical character on the dynamics. The suppression of the drift
resistive instability in the case of sufficiently strong flow shear
is a non--modal process, during which the initial amplitude of the
separate spatial Fourier mode of the perturbed electrostatic
potential decreases with time as $(v'_{0}t)^{-2}$.

In this paper we extend the non-modal treatment of the drift modes
stability onto the plasma flows with a spatially homogeneous
time-dependent velocity shear.  In Section II we consider linear and
weak-nonlinear solutions of the nonlinear Hasegawa-Mima equation for
plasma flow with strong time-dependent velocity shear. In Section
III we discuss the correlation properties of drift waves in plasma
shear flows and their influence on the saturation of the resistive
drift instability. In that section we pay the attention to flows
with a "modest" velocity shear (of the order of or greater than the
instability growth rate, but less than the drift waves frequency).
In this case the stage at which nonlinear processes determine the
saturation of the drift resistive instability occurs before the
development of the linear non-modal suppression of drift
instability. Conclusions are given in Section IV.

\section*{II. Non-modal drift waves in plasma flow with strong
time-dependent velocity shear}

We investigate the temporal evolution of drift modes in
time-dependent shear flow using the Hasegawa--Wakatani
equations\cite{Hasegava}. We use Cartesian coordinates aligned so
that $z$ is in the direction of the mean magnetic field, $y$ is in
the direction of the mean flow, and $x$ is in the direction of the
inhomogeneity of the mean flow and plasma density, assuming that
these coordinates correspond to toroidal $\left(\varphi\right)$,
poloidal $\left(\theta\right)$, and the radial $\left(r\right)$
coordinates, respectively, of the toroidal coordinate system. In
this geometry the Hasegawa--Wakatani system of equations for the
dimensionless density $n = \tilde {n} / n_{e} $ and potential $\phi
= e\varphi / T_{e}$ perturbations ($n_{e} $ is the electron
background density, $T_{e}$ is the electron temperature) is
\cite{Hasegava}
\begin{equation}
\rho _{s}^{2} \left(\frac{\partial}{\partial t} + V_{0}
\left(x,t\right)\frac{\partial}{\partial y}-
\frac{c}{B}\left(\frac{\partial\phi}{\partial
y}\frac{\partial}{\partial x}- \frac{\partial\phi}{\partial
x}\frac{\partial}{\partial y}\right)\right)\nabla ^{2}\phi =
a\frac{\partial ^{2}}{\partial z^{2}}\left(n -\phi \right),\label{1}
\end{equation}
\begin{equation}
 \left(\frac{\partial}{\partial  t}
+ V_{0} \left(x,t\right)\frac{\partial}{\partial y}-
\frac{c}{B}\left(\frac{\partial\phi}{\partial
y}\frac{\partial}{\partial x}- \frac{\partial\phi}{\partial
x}\frac{\partial}{\partial y}\right)\right)n + v_{de}
\frac{\partial\phi }{\partial y} = a\frac{\partial ^{2}}{\partial
z^{2}}\left(n - \phi \right),\label{2}
\end{equation}
$V_{0} \left(x,t\right)$ is the velocity  of the sheared flow, $a =
T_{e} / n_{0} e^{2}\eta _{_{\parallel}}$, $\eta _{_{\parallel}}$ is
the resistivity parallel to the homogeneous magnetic field
$\mathbf{B}_{\parallel}\mathbf{z}$, $\rho _{s}$ is the ion Larmor
radius at electron temperature $T_{e}$, $v_{de}=c T_{e}/eBL_{n}$ is
the diamagnetic drift velocity, $L_{n}^{-1}=-d\ln
n_{0e}\left(x\right)/dx$. We transform Eqs.(\ref{1}), (\ref{2}) to
new spatial variables $\xi,\eta$,
\begin{equation}
t=t,\qquad \xi = x,\qquad \eta=y-\int V_{0}\left(x,t\right)dt
,\qquad z = z.\label{3}
\end{equation}
which are the generalizations of the convective coordinates, used
previously in systems with homogeneously sheared stationary
flows\cite{Mikhailenko-2000}. In these coordinates the linear
convection terms in the system of equations (\ref{1}), (\ref{2}) are
excluded, but the Laplacian operator $\Delta $\ now becomes
time-dependent,
\begin{equation}
\Delta =\frac{\partial ^{2}}{\partial x^{2}}+\frac{\partial
^{2}}{\partial
y^{2}}=\left( \frac{\partial }{%
\partial \xi }-\int \frac{\partial V_{0}\left(\xi, t\right)}
{\partial\xi}dt\frac{\partial }{\partial \eta }\right)\left( \frac{\partial }{%
\partial \xi }-\int \frac{\partial V_{0}\left(\xi, t\right)}
{\partial\xi}dt\frac{\partial }{\partial \eta
}\right)+\frac{\partial ^{2}}{\partial \eta ^{2}}, \label{4}
\end{equation}
leaving us with an initial value problem to solve. This  time
dependence is responsible for the shearing of the waves pattern by
the basic flow in the convective frame of reference.

The analysis of the effect of the flow shear spatial non-uniformity
on the temporal evolution of the electrostatic potential of drift
modes has shown\cite{Mikhailenko-2008}, that for the velocity
profiles without inflection point this effect is subdominant at all
stages of the drift modes evolution. Therefore we consider here the
case of the spatially uniform time-dependent velocity shear, for
which
\begin{equation}
\int \frac{\partial V_{0}\left(\xi, t\right)}
{\partial\xi}dt=v'_{0}\alpha\left(t\right),\label{5}
\end{equation}
where $v'_{0}=const$ is a parameter with dimension of the velocity
shear, $\alpha\left(t\right)$ is a function of time. Introducing
Fourier expansion with respect to $\xi$, $\eta$ and $z$--coordinates
with $k_{\perp}$, $l$ and $k_{z}$ as the wave numbers conjugate
tothe spatial coordinates $\xi$, $\eta$ and $z$, respectively, the
linearized system (\ref{1}), (\ref{2}) with omitted nonlinear
convective terms may be combined into the following equation for the
potential $\phi\left(t, k_{\perp},l, k_{z}\right)$:
\begin{eqnarray}
&\displaystyle
\frac{1}{C}\frac{1}{\left(v'_{0}\right)^{2}}\frac{\partial^{2}}
{\partial t^{2}}\left(\Delta_{\perp c}\phi\right)+\frac{\rho
_{s}^{2}}{v'_{0}}\frac{\partial}{\partial t}\left(\Delta_{\perp
c}\phi\right)-\frac{1}{v'_{0}}\frac{\partial \phi}{\partial
t}-iS\phi=0.\label{6}
\end{eqnarray}
In Eq.(\ref{4}) $C = ak_{z}^{2}/\rho _{s}^{2} l^{2}v '_{0} = T_{e}
k_{z}^{2}/\rho_{s}^{2} l^{2}\:v'_{0} n_{0}e^{2}\eta
_{_{\parallel}}\gg 1$, $S=lv_{de}/v'_{0}$. The Laplacian operator
$\Delta$ is equal to
\begin{eqnarray}
&\displaystyle \Delta\left(\tau,k_{\perp},l\right)=-\left(k_{\perp}
-lv'_{0}\alpha\left(t\right)\right)^{2}-l^{2}. \label{7}
\end{eqnarray}

In variables $\xi$ and $\eta$ the linear solution to Eq.(\ref{6})
has a form
\begin{eqnarray}
&\displaystyle \phi\left(\xi, \eta, t\right)=\int dk_{\perp}\int dl
\phi\left(k_{\perp},l,0\right)g\left(k_{\perp},l,t\right)e^{ik_{\perp}\xi+il\eta},\label{8}
\end{eqnarray}
where $\phi\left(k_{\perp},l,0\right)$ is the initial data and
$g\left(k_{\perp},l,t\right)$ is the linearly unstable solution. It
was shown in Ref.\cite{Mikhailenko-2000} (see, also\cite{
Mikhailenko-2008}) for the case of the spatially homogeneous
time-independent velocity shear, that for $v '_{0}\lesssim \gamma$,
where $\gamma$ is growth rate of the resistive drift instability in
plasma without shear flow, the solution
$g\left(k_{\perp},l,t\right)$ in times $t \lesssim \left(v
'_{0}l\rho_{s}\right)^{-1}$ has an ordinary modal form,
\begin{eqnarray}
&\displaystyle g\left(k_{\perp},l,t\right)=e^{i\omega_{d}t+\gamma
t}\label{9}
\end{eqnarray}
with known frequency of drift wave $\mathrm{Re}\,\omega=\omega_{d}=
Sv'_{0}/\left(1+\rho_{s}^{2}\left(k^{2}_{\perp}+l^{2}\right)\right)$
, and growth rate of the resistive drift instability $
\gamma=\mathrm{Im}\,\omega=\omega_{d}\rho_{s}^{2}\left(k^{2}_{\perp}+l^{2}\right)/ak^{2}_{z}
\left(1+\rho_{s}^{2}\left(k^{2}_{\perp}+l^{2}\right)\right)\simeq
v'_{0}S^{2}/\mathbb{C}$, where
$\mathbb{C}=C/\left(k^{2}_{\perp}+l^{2}\right)$. Using solution
(\ref{9}), which has a form as in plasma without any shear flow,
weak nonlinear theory, which is grounded on the Hasegawa-Wakatani
system (\ref{1}), (\ref{2}) or Hasegawa-Mima
equation\cite{Hasegava-Mima} may be easily developed\cite{Horton}.
This theory will be free from the known problem of the solutions
singularities at critical level points $x=x_{0}$, where phase
velocity of wave becomes equal to flow velocity, i.e.
$\omega-k_{y}v_{0}\left(x_{0}\right)=0$. They emerge, when linear,
as well as weak-nonlinear, solutions are obtained by using spectral
transform in time and variables $x$, $y$ of the laboratory frame of
reference. It is worth to note, that it is unreasonable from the
very beginning to apply the spectral expansion in time in the case
of the time dependent flow velocity.

In the laboratory frame solution (\ref{8}) becomes nonseparable in
space and time and therefore quite different from the normal mode
assumption. In the case of the time dependent velocity shear it has
a form
\begin{eqnarray}
&\displaystyle \phi\left(\mathbf{r}, t\right)=\int dk_{\perp}\int dl
\phi\left(k_{\perp},l,0\right)g\left(k_{\perp},l,t\right)
e^{i(k_{\perp}-lv'_{0}\alpha\left(t\right))x+ily}.\label{10}
\end{eqnarray}
Only in times, for which $\left|v'_{0}\alpha\left(t\right)\right|\ll
1$ (in times $t\ll \left|v'_{0}\right|^{-1}$ in the case of
stationary velocity shear ) solution (\ref{10}) has a normal mode
form. Because of the time dependence
$k_{x}=k_{\perp}-lv'_{0}\alpha\left(t\right)$ of the wave number
component along the flow shear, the modes in the laboratory
framebecome increasingly one-dimensional zonal-like as the perturbed
$E\times B$ velocity tilts more and more closely parallel to y-axis.

The modal solution (\ref{8}),(\ref{9}) follows over times $\alpha(t)
v '_{0}l\rho_{s}\gtrsim 1$ strong non-modal evolution of
$\phi\left(t,k_{\perp},l\right)$, resulted from the enhanced by
shear flow time-dependent dispersion, which does not allow the weak
nonlinear three waves decay processes. For strong flow shear case,
when $lv_{de}\lesssim v '_{0}$, the first term of Eq.(\ref{6}) over
times $t \gtrsim \left(lv_{de}\right)^{-1}$ may be
omitted\cite{Mikhailenko-2008}. In that case Eq.(\ref{6}) reduces to
the Hasegawa-Mima equation\cite{Hasegava-Mima}, which with
convective nonlinearity (omitted in Eq.(\ref{6})) in the case of the
time-dependent velocity shear has a form
\begin{eqnarray}
&\displaystyle \frac{\partial}{\partial
t}\left[\sigma\left(k_{\perp},l,t\right)\phi\left(t,k_{\perp},l\right)\right]
+ilv_{de}\phi\left(\tau,k_{\perp},l\right)=\rho
_{s}^{2}\frac{c}{B}\int dk_{\perp 1}\int dl_{1}\int dk_{\perp 2}\int
dl_{2}\Delta\left(t,k_{2\perp},l_{2}\right)\nonumber
\\  &\displaystyle \times\left(k_{1\perp}l_{2}-k_{2\perp}l_{1}\right)
\phi\left(t,k_{1\perp},l_{1}\right)
\phi\left(t,k_{2\perp},l_{2}\right)\delta\left(k_{\perp}-k_{1\perp}
-k_{2\perp}\right) \delta\left(l-l_{1}-l_{2}\right),\label{11}
\end{eqnarray}
where $\Delta\left(\tau,k_{2\perp},l_{2}\right)$ is determined by
Eq.(\ref{7})) and
\begin{eqnarray} &\displaystyle
\sigma\left(k_{\perp},l,t\right)=
1+\rho_{s}^{2}\left(l^{2}+\left(k_{\perp}-v'_{0}\alpha
\left(t\right)l\right)^{2}\right).\label{12}
\end{eqnarray}
The approximate weak nonlinear solution to Eq.(\ref{11})) for the
potential $\phi$ is easily obtained in the form of the power series
in the amplitude of the initial data $\phi\left(\xi,l,0\right)$,
$\phi\left(\tau,k_{\perp},l\right)=\phi_{0}\left(\tau,k_{\perp},l\right)
+\phi_{1}\left(\tau,k_{\perp},l\right)+...$, where $\phi_{0}$ and
$\phi_{1}$ are linear and square terms with respect to
$\phi\left(\xi,l,0\right)$, with
\begin{eqnarray}
&\displaystyle \phi_{0}\left(k_{\perp},l,
t\right)=\phi\left(k_{\perp},l,0\right)g\left(k_{\perp},l,t\right)
,\label{13}
\end{eqnarray}
and
\begin{eqnarray}
&\displaystyle \phi_{1}\left(k_{\perp},l,
t\right)=\frac{\rho_{s}^{2}c} {B\sigma\left(k_{\perp},l,t\right)}
\int\limits^{t}_{t_{0}} dt_{1}\mathrm\exp\left[-ilv_{de}\int
\limits^{t_{1}}_{t_{0}} dt_{2}
\sigma^{-1}\left(k_{\perp},l,t_{2}\right)\right]\int dk_{\perp
1}\int dl_{1}\int dk_{\perp 2}\int dl_{2}\nonumber
\\  &\displaystyle \times\Delta\left(t_{1},k_{2\perp},l_{2}\right)
\left(k_{1\perp}l_{2}-k_{2\perp}l_{1}\right)
\phi\left(k_{1\perp},l_{1},0\right)\phi\left(k_{2\perp},l_{2},0\right)
g\left(k_{1\perp},l_{1},t\right)
g\left(k_{2\perp},l_{2},t\right)\nonumber
\\  &\displaystyle \times
\delta\left(k_{\perp}-k_{1\perp}-k_{2\perp}\right)
\delta\left(l-l_{1}-l_{2}\right),\label{14}
\end{eqnarray}
where
\begin{eqnarray}
&\displaystyle
g\left(k_{\perp},l,t\right)=\frac{1}{\sigma\left(k_{\perp},l,t\right)}
\mathrm\exp\left[-ilv_{de}\int \limits^{t}_{t_{0}} dt_{1}
\sigma^{-1}\left(k_{\perp},l,t_{1}\right) \right].\label{15}
\end{eqnarray}

It follows from Eqs.(\ref{13}) and (\ref{14}) that over times, for
which  $\left|\alpha(t) v '_{0}l\right|\rho_{s}\gtrsim 1$ the
initial perturbations of the potential with given $l\rho_{s}$ will
be suppressed before the development of the resistive drift
instability, when non-modal evolution begins at times
$t<\gamma^{-1}$. For the particular case of the time-independent
flow shear solution (\ref{10}) was obtained and discussed in
Ref.\cite{Mikhailenko-2000}. For power-like time dependence of the
velocity shear with $\partial
V_{0}/\partial\xi=v'_{0}\alpha_{n}t^{n}$,
$\alpha\left(t\right)=\alpha_{n}t^{n+1}/(n+1)$ and
\begin{eqnarray} &\displaystyle
\sigma\left(k_{\perp},l,t\right)=
1+\rho_{s}^{2}\left(l^{2}+\left(k_{\perp}-v'_{0}l\frac{\alpha_{n}
t^{n+1}}{(n+1)}\right)^{2}\right),\label{16}
\end{eqnarray}
the non-modal development begins at times
$t>\left(v'_{0}\alpha_{n}l\rho_{s}\right)^{-\frac{1}{n+1}}$, at
which linear $\phi_{0}\left(t,k_{\perp},l\right)$ and weakly
nonlinear $\phi_{1}\left(t,k_{\perp},l\right)$ parts of the
potential rapidly decay with time. The fluctuation spectrum for long
time will be dominated by Fourier components with sufficiently small
values of wave number component $l$ for which condition
$\left|\alpha(t) v '_{0}l\right|\rho_{s}\gtrsim 1$ is not met during
time considered.

\section*{III.Renormalized hydrodynamic theory for drift modes in plasma shear flow}

With variables $\xi, \eta$,  Eqs. (\ref{1}) and (\ref{2}) have a
form
\begin{equation}
\rho _{s}^{2} \left(\frac{\partial}{\partial t}-
\frac{c}{B}\left(\frac{\partial\phi}{\partial\eta}\frac{\partial}{\partial\xi}-
\frac{\partial\phi}{\partial\xi}\frac{\partial}{\partial\eta}\right)\right)\nabla
^{2}\phi = a\frac{\partial ^{2}}{\partial z^{2}}\left(n -\phi
\right),\label{17}
\end{equation}
\begin{equation}
 \left(\frac{\partial}{\partial  t}-
\frac{c}{B}\left(\frac{\partial\phi}{\partial\eta}\frac{\partial}{\partial\xi}-
\frac{\partial\phi}{\partial\xi}\frac{\partial}{\partial\eta}\right)\right)n
+ v_{de} \frac{\partial\phi }{\partial y} = a\frac{\partial
^{2}}{\partial z^{2}}\left(n - \phi \right).\label{18}
\end{equation}
The above calculations show that with variables $\xi$ and $\eta$ we
exclude from Eqs.(\ref{1}) and (\ref{2}) the spatial inhomogeneity
originated from shear flow. That gives an opportunity to investigate
in linear approximation the temporal evolution of the separate
spatial Fourier mode of the electrostatic potential $\phi$ with
definite wave numbers $k_{\perp}$ and $l$.  It is interesting to
note that transformation (\ref{3}) conserves the $E\times B$
convective nonlinear derivative in Eqs.(\ref{1}) and (\ref{2}) in
the form similar to one in a plasma without any flows. The flow
shear contains only in time dependent Laplacian operator and become
apparent in linear non-modal effect of the enhanced dispersion
during times $\left|\alpha(t) v '_{0}l\right|\rho_{s}\gtrsim 1$ (
or, at the case of stationary velocity shear, when
$t>\left(v'_{0}l\rho_{s}\right)^{-1}$) after the stage of the
ordinary modal evolution\cite{Mikhailenko-2000}. With new variables
$\xi_{1}, \eta_{1}$, which are determined by the nonlinear relations
\begin{eqnarray}
&\displaystyle \xi_{1}=\xi
+\frac{c}{B}\int\limits_{t_{0}}^{t}\frac{\partial\phi}{\partial\eta}dt_{1},
\qquad\qquad \eta_{1}=\eta
-\frac{c}{B}\int\limits_{t_{0}}^{t}\frac{\partial\phi}{\partial\xi}dt_{1}\label{19},
\end{eqnarray}
the convective nonlinearity in Eqs.(\ref{17}) and (\ref{18}) becomes
of the higher order with respect to the potential $\phi$,
\begin{eqnarray*}
&\displaystyle -
\frac{c}{B}\left(\frac{\partial\phi}{\partial\eta}\frac{\partial}{\partial\xi}-
\frac{\partial\phi}{\partial\xi}\frac{\partial}{\partial\eta}\right)=\nonumber
\\  &\displaystyle-\left(\frac{c}{B}\right)^{2}\left[\left(\frac{\partial\phi\left(t\right)}
{\partial\eta}\int\limits_{t_{0}}^{t}
dt_{1}\frac{\partial^{2}\phi\left(t_{1}\right)}{\partial\eta\partial\xi}-
\frac{\partial\phi\left(t\right)}{\partial\xi}\int\limits_{t_{0}}^{t}
dt_{1}\frac{\partial^{2}\phi\left(t_{1}\right)}{\partial\eta^{2}}
\right)\frac{\partial}{\partial\xi_{1}}\right.\nonumber
\\  &\displaystyle\left.+\left(\frac{\partial\phi\left(t\right)}{\partial\xi_{1}}\int\limits_{t_{0}}^{t}
dt_{1}\frac{\partial^{2}\phi\left(t_{1}\right)}{\partial\eta\partial\xi}-
\frac{\partial\phi\left(t\right)}{\partial\eta}\int\limits_{t_{0}}^{t}
dt_{1}\frac{\partial^{2}\phi\left(t_{1}\right)}{\partial\xi^{2}}
\right)\frac{\partial}{\partial\eta_{1}}\right]
\end{eqnarray*}
\begin{eqnarray}
&\displaystyle-\left(\frac{c}{B}\right)^{3}
\left[\int\limits_{t_{0}}^{t}
dt_{1}\frac{\partial^{2}\phi\left(t_{1}\right)}
{\partial\xi^{2}}\int\limits_{t_{0}}^{t}
dt_{1}\frac{\partial^{2}\phi\left(t_{1}\right)}
{\partial\eta^{2}}-\left(\int\limits_{t_{0}}^{t}
dt_{1}\frac{\partial^{2}\phi\left(t_{1}\right)}
{\partial\eta\partial\xi}\right)^{2}\right]\nonumber
\\  &\displaystyle\times\left(\frac{\partial\phi\left(t\right)}
{\partial\eta_{1}}\frac{\partial}{\partial\xi_{1}}-
\frac{\partial\phi\left(t\right)}{\partial\xi_{1}}
\frac{\partial}{\partial\eta_{1}}\right).\label{20}
\end{eqnarray}
Omitting such nonlinearity, as well as small nonlinearity of the
second order in the Laplacian, resulted from the transformation to
nonlinearly determined variables $\xi_{1}, \eta_{1}$, we come to
linear equation (\ref{6}) with solution (\ref{8}), where wave
numbers $k_{\perp},l$ are conjugate there to coordinates $\xi_{1}$,
$\eta_{1}$ respectively. With variables $\xi$ and $\eta$ this
solution has a form
\begin{eqnarray}
&\displaystyle \phi\left(\xi, \eta, t\right)=\int dk_{\perp}\int dl
\phi\left(k_{\perp}, l, 0\right)g\left(k_{\perp}, l,
t_{1}\right)e^{ik_{\perp}\xi+il\eta-ik_{\perp}\widetilde{\xi}
\left(t_{1}\right)-il\widetilde{\eta}\left(t_{1}\right)},\label{21}
\end{eqnarray}
where
\begin{eqnarray}
&\displaystyle
\widetilde{\xi}\left(t\right)=-\frac{c}{B}\int\limits_{t_{0}}^{t}\frac{\partial\phi
\left(t_{1}\right)}{\partial\eta}dt_{1}, \qquad\qquad
\widetilde{\eta}\left(t\right)=\frac{c}{B}\int\limits_{t_{0}}^{t}\frac{\partial\phi
\left(t_{1}\right)}{\partial\xi}dt_{1}.\label{22}
\end{eqnarray}
Eq.(\ref{21}) is in fact a nonlinear integral equation for potential
$\phi$, in which the effect of the total Fourier spectrum on any
separate Fourier harmonic is accounted for. The functions
$\widetilde{\xi} \left(t\right)$ and
$\widetilde{\eta}\left(t\right)$ in the exponential of Eq.(\ref{21})
involve through Eq.(\ref{22}) integrals of $\phi$, which in turn,
involve in their exponentials the integrals (\ref{22}) and so on.
This form of solution, however, appears very useful for the analysis
of the correlation properties of the nonlinear solutions to
Hasegawa-Wakatani system and for the development of the approximate
renormalized solutions to Hasegawa-Wakatani system, which accounted
for the effect of the turbulent motions of plasma on the saturation
of the drift-resistive instability. Consider first the dispersion
tensor of random displacements $\widetilde{\xi}\left(t\right)$ and
$\widetilde{\eta}\left(t\right)$ of the plasma resulted from its
motion in such turbulence. The variances of these displacements are
determined by the relations
\begin{eqnarray}
&\displaystyle K_{\xi\xi}\left(t,
t_{0}\right)=K_{\xi\xi}\left(t\right)=\left\langle\left(\widetilde{\xi}
\left(t\right)\right)^{2}\right\rangle=
\frac{c^{2}}{B^{2}}\int\limits_{t_{0}}^{t}
dt_{1}\int\limits_{t_{0}}^{t} dt_{2}\left\langle\frac{\partial\phi
\left(t_{1}\right)}{\partial\eta}\frac{\partial\phi
\left(t_{2}\right) }{\partial\eta}\right\rangle,\label{23}
\end{eqnarray}
\begin{eqnarray}
&\displaystyle K_{\xi\eta}\left(t,
t_{0}\right)=K_{\xi\eta}\left(t\right)=\left\langle\left(\widetilde{\xi}\left(t\right)
\widetilde{\eta}\left(t\right)\right)\right\rangle=
-\frac{c^{2}}{B^{2}}\int\limits_{t_{0}}^{t}
dt_{1}\int\limits_{t_{0}}^{t} dt_{2}\left\langle\frac{\partial\phi
\left(t_{1}\right)}{\partial\eta}\frac{\partial\phi
\left(t_{2}\right) }{\partial\xi}\right\rangle,\label{24}
\end{eqnarray}
\begin{eqnarray}
&\displaystyle K_{\eta\eta}\left(t,
t_{0}\right)=K_{\eta\eta}\left(t\right)=\left\langle\left(\widetilde{\eta}
\left(t\right)\right)^{2}\right\rangle=
\frac{c^{2}}{B^{2}}\int\limits_{t_{0}}^{t}
dt_{1}\int\limits_{t_{0}}^{t} dt_{2}\left\langle\frac{\partial\phi
\left(t_{1}\right)}{\partial\xi}\frac{\partial\phi
\left(t_{2}\right) }{\partial\xi}\right\rangle,\label{25}
\end{eqnarray}
where brackets $\left\langle...\right\rangle$ mean an average over
ensemble of random values of $\phi\left(k_{\perp},l,0\right)$. Using
in Eqs.(\ref{23})-(\ref{25}) solution (\ref{21}) and the
Corrsin's\cite{Corrsin} independence hypothesis,
\begin{eqnarray}
&\displaystyle \left\langle
\phi\left(k_{1\perp},l_{1},t_{1}\right)\phi\left(k_{2\perp},l_{2},t_{2}\right)
e^{ik_{\perp}\xi+il\eta-ik_{\perp}\widetilde{\xi}
\left(t_{1}\right)-il\widetilde{\eta}\left(t_{1}\right)}
\right\rangle=\left\langle\phi\left(k_{1\perp},l_{1},
0\right)\phi\left(k_{2\perp},l_{2},0\right)\right\rangle\nonumber
\\  &\displaystyle \times
g\left(k_{1\perp},l_{1},t_{1}\right)g\left(k_{2\perp},l_{2},t_{2}\right)
\left\langle
e^{ik_{1\perp}\left(\widetilde{\xi}\left(t_{1}\right)-\widetilde{\xi}\left(t_{2}\right)\right)
+il_{1}\left(\widetilde{\eta}\left(t_{1}\right)-\widetilde{\eta}\left(t_{2}\right)\right)}\right\rangle
\delta\left(k_{1\perp}+k_{2\perp}\right)\delta\left(l_{1}+l_{2}\right),\label{26}
\end{eqnarray}
we find for the components $K_{\xi\xi}\left(t\right)$ ,
$K_{\xi\eta}\left(t\right)$ and $K_{\eta\eta}\left(t\right)$ of the
dispersion tensor the expressions
\begin{eqnarray}
&\displaystyle K_{\xi\xi}\left(t\right)=
\frac{c^{2}}{B^{2}}\int\limits_{t_{0}}^{t}
dt_{1}\int\limits_{t_{0}}^{t} dt_{2}\int dk_{\perp}\int dl
\left|\phi\left(k_{\perp},l,0\right)\right|^{2}l^{2}g\left(k_{\perp},l,t_{1}\right)
g\left(-k_{\perp},-l,t_{2}\right)\nonumber
\\  &\displaystyle \times \left\langle
e^{ik_{1\perp}\left(\widetilde{\xi}\left(t_{1}\right)-\widetilde{\xi}\left(t_{2}\right)\right)
+il_{1}\left(\widetilde{\eta}\left(t_{1}\right)-\widetilde{\eta}
\left(t_{2}\right)\right)}\right\rangle,\label{27}
\end{eqnarray}
\begin{eqnarray}
&\displaystyle K_{\xi\eta}\left(t\right)=
-\frac{c^{2}}{B^{2}}\int\limits_{t_{0}}^{t}
dt_{1}\int\limits_{t_{0}}^{t} dt_{2}\int dk_{\perp}\int dl
\left|\phi\left(k_{\perp},l,0\right)\right|^{2}k_{\perp}l
g\left(k_{\perp},l,t_{1}\right)
g\left(-k_{\perp},-l,t_{2}\right)\nonumber
\\  &\displaystyle \times \left\langle
e^{ik_{1\perp}\left(\widetilde{\xi}\left(t_{1}\right)-\widetilde{\xi}\left(t_{2}\right)\right)
+il_{1}\left(\widetilde{\eta}\left(t_{1}\right)-\widetilde{\eta}
\left(t_{2}\right)\right)}\right\rangle,\label{28}
\end{eqnarray}
\begin{eqnarray}
&\displaystyle K_{\eta\eta}\left(t\right)=
\frac{c^{2}}{B^{2}}\int\limits_{t_{0}}^{t}
dt_{1}\int\limits_{t_{0}}^{t} dt_{2}\int dk_{\perp}\int dl
\left|\phi\left(k_{\perp},l,0\right)\right|^{2}l^{2}g\left(k_{\perp},l,t_{1}\right)
g\left(-k_{\perp},-l,t_{2}\right)\nonumber
\\  &\displaystyle \times \left\langle
e^{ik_{1\perp}\left(\widetilde{\xi}\left(t_{1}\right)-\widetilde{\xi}\left(t_{2}\right)\right)
+il_{1}\left(\widetilde{\eta}\left(t_{1}\right)-\widetilde{\eta}
\left(t_{2}\right)\right)}\right\rangle.\label{29}
\end{eqnarray}
During times $\left|\alpha(t) v '_{0}l\right|\rho_{s}\lesssim 1$,
the solution $g$ to the Wasegawa-Wakatani system is of a modal form
(\ref{9}). Assuming that the displacements
$\widetilde{\xi}\left(t\right)$, $\widetilde{\eta}\left(t\right)$
obey the Gaussian statistics with mean zero, we find in this case
the following relation for $K_{\xi\xi}\left(t\right)$:
\begin{eqnarray}
&\displaystyle K_{\xi\xi}\left(t\right)=
\frac{c^{2}}{B^{2}}\int\limits_{t_{0}}^{t}
dt_{1}\int\limits_{t_{0}}^{t} dt_{2}\int dk_{\perp}\int dl
\left|\phi\left(k_{\perp},l,0\right)\right|^{2}l^{2}\exp\left(\gamma\left(t_{1}+t_{2}\right)
+i\omega_{d}\left(t_{1}-t_{2}\right)\right)\nonumber
\\  &\displaystyle \times \exp\left[-\frac{1}{2}k^{2}_{\perp}
K_{\xi\xi}\left(t_{1},t_{2}\right) - k_{\perp}l
K_{\xi\eta}\left(t_{1},t_{2}\right)-\frac{1}{2}l^{2}
K_{\eta\eta}\left(t_{1},t_{2}\right)\right]. \label{30}
\end{eqnarray}
where
\begin{eqnarray}
&\displaystyle K_{\xi\xi}\left(t_{1},t_{2}\right)=
\left\langle\left(\widetilde{\xi}\left(t_{1}\right)-\widetilde{\xi}
\left(t_{2}\right)\right)^{2}\right\rangle, \nonumber
\\  &\displaystyle
K_{\xi\eta}\left(t_{1},t_{2}\right)=\left\langle\left(\widetilde{\xi}
\left(t_{1}\right)-\widetilde{\xi}\left(t_{2}\right)\right)
\left(\widetilde{\eta}\left(t_{1}\right)
-\widetilde{\eta}\left(t_{2}\right)\right)\right\rangle,\nonumber
\\  &\displaystyle
K_{\eta\eta}\left(t_{1},t_{2}\right)=\left\langle\left(\widetilde{\eta}
\left(t_{1}\right)-\widetilde{\eta}\left(t_{2}\right)\right)^{2}\right\rangle.
\label{31}
\end{eqnarray}
With time variables $\tau=t_{1}-t_{2}$ and
$\widehat{t}=\left(t_{1}+t_{2}\right)/2$, Eq.(\ref{30}) becomes
\begin{eqnarray}
&\displaystyle
K_{\xi\xi}\left(t\right)=\frac{c^{2}}{B^{2}}\left[\int\limits_{-t}^{0}
d\tau\int\limits_{-\frac{\tau}{2}}^{t+\frac{\tau}{2}}
d\widehat{t}+\int\limits_{0}^{t}
d\tau\int\limits_{\frac{\tau}{2}}^{t-\frac{\tau}{2}}
d\widehat{t}\right]\int dk_{\perp}\int dl
\left|\phi\left(k_{\perp},l,0\right)\right|^{2}l^{2}\exp\left(2\gamma\left(k_{\perp},l\right)\widehat{t}
+i\omega_{d}\left(k_{\perp},l\right)\tau\right)\nonumber
\\  &\displaystyle \times \exp\left[-\frac{c^{2}}{2B^{2}}\int\limits_{-\tau}^{0}
d\tau_{1}\int\limits_{\widehat{t}-\frac{1}{2}\left(\tau+\tau_{1}\right)}^
{\widehat{t}+\frac{1}{2}\left(\tau+\tau_{1}\right)}
d\widehat{t_{1}}\int dk_{1\perp}\int dl_{1}
\left|\phi\left(k_{1\perp},l_{1},0\right)\right|^{2}\left|\left[\mathbf{k}_{\perp}\times
\mathbf{k}_{1\perp}\right]\right|^{2}\right.\nonumber
\\  &\displaystyle \left.\times\exp\left(2\gamma\left(k_{1\perp},l_{1}\,\right)
\widehat{t_{1}}+i\omega_{d}\left(k_{1\perp},l_{1}\,\right)\tau_{1}-\frac{1}{2}
\left\langle\left(\mathbf{k}_{1\perp}\cdot\widetilde{\mathbf{r}}\left(\widehat{t_{1}}+\frac{\tau_{1}}{2}\right)-
\mathbf{k}_{1\perp}\cdot\widetilde{\mathbf{r}}\left(\widehat{t_{1}}
-\frac{\tau_{1}}{2}\right)\right)^{2}\right\rangle\right)
\right.\nonumber
\\  &\displaystyle \left.-\frac{c^{2}}{2B^{2}}\int\limits_{0}^{\tau}
d\tau_{1}\int\limits_{\widehat{t}-\frac{1}{2}\left(\tau-\tau_{1}\right)}^{\widehat{t}
+\frac{1}{2}\left(\tau-\tau_{1}\right)}
d\widehat{t_{1}}\int dk_{1\perp}\int dl_{1}
\left|\phi\left(k_{1\perp},l_{1},0\right)\right|^{2}\left|\left[\mathbf{k}_{\perp}\times
\mathbf{k}_{1\perp}\right]\right|^{2}\right.\nonumber
\\  &\displaystyle \left.\times\exp\left(2\gamma\left(k_{1\perp},l_{1}\,\right)
\widehat{t_{1}}+i\omega_{d}\left(k_{1\perp},l_{1}\,\right)\tau_{1}-\frac{1}{2}
\left\langle\left(\mathbf{k}_{1\perp}\cdot\widetilde{\mathbf{r}}\left(\widehat{t_{1}}+\frac{\tau_{1}}{2}\right)-
\mathbf{k}_{1\perp}\cdot\widetilde{\mathbf{r}}\left(\widehat{t_{1}}
-\frac{\tau_{1}}{2}\right)\right)^{2}\right\rangle\right)\right],
\label{32}
\end{eqnarray}
where $\widetilde{\mathbf{r}}=(\widetilde{\xi}, \widetilde{\eta})$
and $\mathbf{k}_{\perp}=\left(k_{\perp},l\right)$. The similar
relations (see Eqs.(\ref{27})-(\ref{29}) above) may be obtained for
$K_{\eta\eta}$ and $K_{\xi\eta}$, changing $l^{2}$ with
$k_{\perp}^{2}$ for $K_{\eta\eta}$ and with $-k_{\perp}l$ in the
integrand for $K_{\xi\eta}$, respectively. Eq.(\ref{32}) contains
two time scales. The fast time scale, determined by
$\omega_{d}^{-1}$, is the correlation time and the slow time scale,
determined by $\gamma^{-1}$ may be considered as the relaxation
time. In the case, when $\gamma^{-1}\gg\omega_{d}^{-1}$,
Eq.(\ref{32}) may be substantially simplified. As it follows from
(\ref{32}), the integrals over slow time $\widehat{t_{1}}$ and
infinite sequence of integrals, which appear in
$\widetilde{\mathbf{r}}\left(\widehat{t_{1}}
\pm\frac{\tau_{1}}{2}\right)$ over slow time scales, are calculated
over narrow time intervals of the order of $\tau\sim
\tau_{correlation}\sim\omega_{d}^{-1}$ over which the variation of
the integrands that occurs from the slow time is negligible. In such
case the integrals over $\widehat{t_{1}}$ may be approximately
calculated as
\begin{eqnarray}
&\displaystyle\int\limits_{\widehat{t}-\frac{1}{2}\left(\tau\pm\tau_{1}\right)}
^{\widehat{t}+\frac{1}{2}\left(\tau\pm\tau_{1}\right)}
d\widehat{t_{1}}f\left(\widehat{t_{1}}\right)\approx
\left(\tau\pm\tau_{1}\right)f\left(\widehat{t}\right).\label{33}
\end{eqnarray}
With quasi-Markovian approximation,
\begin{eqnarray}
&\displaystyle \frac{1}{2}
\left\langle\left(\mathbf{k}_{1\perp}\cdot\widetilde{\mathbf{r}}\left(\widehat{t_{1}}+\frac{\tau_{1}}{2}\right)-
\mathbf{k}_{1\perp}\cdot\widetilde{\mathbf{r}}\left(\widehat{t_{1}}
-\frac{\tau_{1}}{2}\right)\right)^{2}\right\rangle\approx \tau_{1}
C\left(k_{1\perp},l_{1},\widehat{t}_{1}\,\right)\label{34}
\end{eqnarray}
which, in fact, resulted from Eq.(\ref{33}), we obtain the integral
equation for the function $C\left(k_{\perp},l, \widehat{t}\right)$:
\begin{eqnarray}
&\displaystyle \frac{1}{2}\left(k_{\perp}^{2}K_{\xi\xi}\left(t_{1},
t_{2}\right)+k_{\perp}lK_{\xi\eta}\left(t_{1},
t_{2}\right)+l^{2}K_{\eta\eta}\left(t_{1},
t_{2}\right)\right)\nonumber \\
&\displaystyle=\frac{1}{2}\left(k_{\perp}^{2}K_{\xi\xi}\left(\widehat{t}+\frac{\tau}{2},
\widehat{t}-\frac{\tau}{2}\right)+k_{\perp}lK_{\xi\eta}\left(\widehat{t}+\frac{\tau}{2},
\widehat{t}-\frac{\tau}{2}\right)+l^{2}K_{\eta\eta}\left(\widehat{t}+\frac{\tau}{2},
\widehat{t}-\frac{\tau}{2}\right)\right)\nonumber \\
&\displaystyle
=C\left(k_{\perp}, l, \widehat{t}\right)\tau\nonumber \\
&\displaystyle=\tau\frac{c^{2}}{B^{2}}\int dk_{1\perp}\int dl_{1}
\left|\phi\left(k_{1\perp},l_{1},\widehat{t}\,\,\right)\right|^{2}\left|\left[\mathbf{k}_{\perp}\times
\mathbf{k}_{1\perp}\right]\right|^{2}\int\limits_{0}^{\infty}
d\tau_{1}e^{\left(i\omega_{d}\left(k_{1\perp},l_{1}\,\right)-C\left(k_{1\perp},l_{1},\widehat{t}\,\right)
\right)\tau_{1}}\nonumber \\
&\displaystyle\approx\tau\frac{c^{2}}{B^{2}}\int dk_{1\perp}\int
dl_{1}
\left|\phi\left(k_{1\perp},l_{1},\widehat{t}\,\,\right)\right|^{2}\left|\left[\mathbf{k}_{\perp}\times
\mathbf{k}_{1\perp}\right]\right|^{2}\frac{C\left(k_{1\perp},l_{1},\widehat{t}\,\right)}
{\omega_{d}^{2}\left(k_{1\perp},l_{1},\,\right)},\label{35}
\end{eqnarray}
where
$\left|\phi\left(k_{1\perp},l_{1},\widehat{t}\,\,\right)\right|^{2}
=\left|\phi\left(k_{1\perp},l_{1},0\,\,\right)\right|^{2}
e^{2\gamma\left(k_{1\perp}, l_{1}\right)\widehat{t}}$. Then
$K_{\xi\xi}\left(t\right)$ becomes equal to
\begin{eqnarray}
&\displaystyle
K_{\xi\xi}\left(t\right)=\frac{c^{2}}{B^{2}}\left[\int\limits_{-t}^{0}
d\tau\int\limits_{-\frac{\tau}{2}}^{t+\frac{\tau}{2}}
d\widehat{t}+\int\limits_{0}^{t}
d\tau\int\limits_{\frac{\tau}{2}}^{t-\frac{\tau}{2}}
d\widehat{t}\right]\int dk_{\perp}\int dl
\left|\phi\left(k_{\perp},l,0\right)\right|^{2}l^{2}e^{2\gamma\left(k_{\perp},l\,\right)\widehat{t}
+i\omega_{d}\left(k_{\perp},l\,\right)\tau}\nonumber
\\  &\displaystyle \times\exp\left[-\tau\frac{c^{2}}{B^{2}}\int dk_{1\perp}\int dl_{1}
\left|\phi\left(k_{1\perp},l_{1},0\right)\right|^{2}e^{2\gamma\left(k_{1\perp},l_{1}\,\right)\widehat{t}}
\left|\left[\mathbf{k}_{\perp}\times
\mathbf{k}_{1\perp}\right]\right|^{2}\frac{C\left(k_{1\perp},l_{1},\widehat{t}\,\right)}
{\omega_{d}^{2}\left(k_{1\perp},l_{1},\,\right)}\right]\nonumber
\\  &\displaystyle \approx \frac{2c^{2}}{B^{2}}\int dk_{\perp}\int
dl l^{2}\int \limits_{0}^{t} d\widehat{t}
\left|\phi\left(k_{\perp},l,\widehat{t}\,\,\right)\right|^{2}
\frac{C\left(k_{\perp},l,\widehat{t}\,\right)}
{\omega_{d}^{2}\left(k_{\perp},l\right)} .\label{36}
\end{eqnarray}
With Eq.(\ref{36}) and similar equations for
$K_{\xi\eta}\left(t\right)$ and $K_{\eta\eta}\left(t\right)$ we
obtain a more general equation,
\begin{eqnarray}
&\displaystyle k^{2}_{\perp}K_{\xi\xi}\left(t\right)+
2k_{\perp}lK_{\xi\eta}\left(t\right)
+l^{2}K_{\eta\eta}\left(t\right)\nonumber \\
&\displaystyle =\frac{c^{2}}{B^{2}}\int dk_{1\perp}\int dl_{1}\int
\limits_{0}^{t} d\widehat{t}
\left|\phi\left(k_{1\perp},l_{1},\widehat{t}\,\,\right)\right|^{2}\left|\left[\mathbf{k}_{\perp}\times
\mathbf{k}_{1\perp}\right]\right|^{2}\frac{C\left(k_{1\perp},l_{1},\widehat{t}\,\right)}
{\omega_{d}^{2}\left(k_{1\perp},l_{1}\right)} =2\int \limits_{0}^{t}
d\widehat{t}C\left(k_{\perp},l,\widehat{t}\,\right).\label{37}
\end{eqnarray}
With Eq.(\ref{37}) we obtain the renormalized form of the potential
(\ref{21}), in which the average effect of the random convection is
accounted for,
\begin{eqnarray}
&\displaystyle \phi\left(\xi, \eta, t\right)=\int dk_{\perp}\int dl
\phi\left(k_{\perp},l,0\right)e^{i\omega_{d}t+\gamma t-\int
\limits_{0}^{t}
d\widehat{t}C\left(k_{\perp},\,l,\,\widehat{t}\,\right)+ik_{\perp}\xi+il\eta}.\label{38}
\end{eqnarray}
The saturation of the instability occurs when $\partial
\left(\phi\left(\xi, \eta, t\right)\right)^{2}/\partial t=0 $, i.e.
when
\begin{eqnarray}
&\displaystyle\gamma\left(k_{\perp},l\right)=C\left(k_{\perp},l,
t\right)\nonumber \\
&\displaystyle=\frac{c^{2}}{B^{2}}\int dk_{1\perp}\int dl_{1}
\left|\phi\left(k_{1\perp},l_{1},t\right)\right|^{2}\left|\left[\mathbf{k}_{\perp}\times
\mathbf{k}_{1\perp}\right]\right|^{2}\frac{C\left(k_{1\perp},\,l_{1},\,t\,\right)}
{\omega_{d}^{2}\left(k_{1\perp},l_{1}\right)}.\label{39}
\end{eqnarray}
From the double Eq.(\ref{39}) we obtain the equation, which
determines the level of the instability saturation
\begin{eqnarray}
&\displaystyle\gamma\left(k_{\perp},l\right)=\frac{c^{2}}{B^{2}}\int
dk_{1\perp}\int dl_{1}
\left|\phi\left(k_{1\perp},l_{1},t\right)\right|^{2}\left|\left[\mathbf{k}_{\perp}\times
\mathbf{k}_{1\perp}\right]\right|^{2}\frac{\gamma\left(k_{1\perp},l_{1}\right)}
{\omega_{d}^{2}\left(k_{1\perp},l_{1}\right)}.\label{40}
\end{eqnarray}
The sought-for value in Eq.(\ref{40}) is a time $t_{sat}$ at which
the balance of the linear growth and nonlinear damping occurs for
given initial disturbance $\phi\left(k_{1\perp},l_{1},0\right)$ and
dispersion. With obtained $t_{sat}$ the saturation level will be
equal to $\left|\phi\left(t_{sat}\right)\right|^{2}\simeq\int
dk_{1\perp}\int
dl_{1}\left|\phi\left(k_{1\perp},l_{1},0\right)\right|^{2}
e^{2\gamma\left(k_{1\perp},\,\,l_{1}\,\right)\, t_{sat}}$. Also, the
well known order of value estimate \cite{Horton} for the potential
$\phi$ in the saturation state is obtained easily from
Eq.(\ref{40}),
\begin{eqnarray}
&\displaystyle \frac{e\phi}{T_{e}}\sim \frac{1}{k_{\perp}
L_{n}}.\label{41}
\end{eqnarray}

Obtained  results show that the nonlinearity of the
Hasegawa-Wakatani system of equations in variables $\xi$ and $\eta$,
with which frequency and growth rate are determined without
spatially inhomogeneous Doppler shift and wave number is time
independent, does not display any effects of the enhanced
decorrelations provided by flow shear. In the laboratory frame of
reference such spatial Fourier modes are observed as a sheared modes
with time dependent component of the wave number $k_{x}=
k_{\perp}-lv'_{0}\alpha\left(t\right)$ directed along the velocity
shear,
\begin{eqnarray}
&\displaystyle \phi\left(\mathbf{r}, t\right)= \int dk_{\perp}\int
dl \phi\left(k_{\perp}, l,
0\right)e^{i(k_{\perp}-lv'_{0}\alpha\left(t\right))x+ily+i\omega_{d}t+\gamma
t-ik_{\perp}\widetilde{\xi}
\left(t_{1}\right)-il\widetilde{\eta}\left(t_{1}\right)}.\label{42}
\end{eqnarray}
To make the next analysis simpler we consider in what follows the
case of the spatially homogeneous time-independent velocity shear,
$\alpha\left(t\right)=t$. The displacements
$\widetilde{\xi}\left(t\right)$ and $\widetilde{\eta}\left(t\right)$
are observed in the laboratory frame as the displacements
$\widetilde{x}\left(t\right)$ and $\widetilde{y}\left(t\right)$
which are equal to
\begin{eqnarray}
&\displaystyle
\widetilde{x}\left(t\right)=\widetilde{\xi}\left(t\right)=-\frac{c}{B}
\int\limits^{t}_{t_{0}}\frac{\partial\phi}{\partial y}
dt_{1}\nonumber
\\  &\displaystyle =-i\frac{c}{B} \int\limits^{t}_{t_{0}}dt_{1}\int
dk_{\perp}\int dl\phi\left(k_{\perp},l,0\right)l
e^{i\omega_{d}t_{1}+\gamma
t_{1}+i(k_{\perp}-lv'_{0}t_{1})x+ily-ik_{\perp}\widetilde{\xi}
\left(t_{1}\right)-il\widetilde{\eta}\left(t_{1}\right)},\label{43}
\end{eqnarray}
and
\begin{eqnarray}
&\displaystyle
\widetilde{y}\left(t\right)=\int\limits^{t}_{t_{0}}\widetilde{v_{y}}\left(t_{1}\right)dt_{1}
=\frac{c}{B}\int\limits^{t}_{t_{0}}dt_{1}\frac{\partial\phi}{\partial
x}+ \int\limits^{t}_{t_{0}}dt_{1}\frac{\partial V_{0}\left(x,
t_{1}\right)} {\partial x} \widetilde{x}\left(t_{1}\right)\nonumber
\\  &\displaystyle=i\frac{c}{B}\int
dk_{\perp}\int dl\int\limits^{t}_{t_{0}}dt_{1}
\phi\left(k_{\perp},l,0\right)\left(k_{\perp}-lv'_{0}\left(t-t_{1}\right)\right)\nonumber
\\  &\displaystyle\times
e^{i\omega_{d}t_{1}+\gamma
t_{1}+i(k_{\perp}-lv'_{0}t_{1})x+ily-ik_{\perp}\widetilde{\xi}
\left(t_{1}\right)-il\widetilde{\eta}\left(t_{1}\right)}.\label{44}
\end{eqnarray}

Applying the procedure presented above to the calculations the
variants of the plasma displacements $K_{xx}\left(t_{1},
t_{2}\right)$ in variables $x$, $y$ we find that $K_{xx}\left(t_{1},
t_{2}\right)$ has the same form as $K_{\xi\xi}\left(t_{1},
t_{2}\right)$ in Eq.(\ref{36}) in which, however, $\omega_{d}$ is
changed by $\omega_{d}-lv'_{0}x$, assuming that
$\omega_{d}-lv'_{0}x\sim \omega_{d}$, and
$C\left(k_{1\perp},\,l_{1},\,t\,\right)$ is changed by
$C_{1}\left(k_{1\perp},\,l_{1},\,t, x\right)$, determined by
Eq.(\ref{39}), in which the replacement $\omega_{d}$ by
$\omega_{d}-lv'_{0}x\sim \omega_{d}$ has to be done. The correlation
$K_{yy}\left(t\right)$ is
\begin{eqnarray}
&\displaystyle
K_{yy}\left(t\right)=\frac{c^{2}}{B^{2}}\left[\int\limits_{-t}^{0}
d\tau\int\limits_{-\frac{\tau}{2}}^{t+\frac{\tau}{2}}
d\widehat{t}+\int\limits_{0}^{t}
d\tau\int\limits_{\frac{\tau}{2}}^{t-\frac{\tau}{2}}
d\widehat{t}\right]\int dk_{\perp}\int dl
\left|\phi\left(k_{\perp},l,\widehat{t}\right)\right|^{2}
e^{i\left(\omega_{d}\left(k_{\perp},l\,\right)-lv'_{0}x\right)\tau-C_{1}\left|\tau\right|}\nonumber
\\  &\displaystyle \times\left(k_{\perp}-lv'_{0}\left(t-\widehat{t}-\frac{\tau}{2}\right)\right)
\left(k_{\perp}-lv'_{0}\left(t-\widehat{t}+\frac{\tau}{2}\right)\right)\nonumber
\\  &\displaystyle \approx \frac{2c^{2}}{B^{2}}\int dk_{\perp}\int
dl \int \limits_{0}^{t} d\widehat{t}
\left|\phi\left(k_{\perp},l,\widehat{t}\,\,\right)\right|^{2}
\left(k_{\perp}-lv'_{0}\left(t-\widehat{t}\right)\right)^{2}
\frac{C_{1}\left(k_{\perp},l,\widehat{t}, x\right)}
{\left(\omega_{d}\left(k_{\perp},l\right)-lv'_{0}x\right)^{2}}
.\label{45}
\end{eqnarray}
The calculation of the integral over $\widehat{t}$ needs the
knowledge of the $C_{1}\left(\widehat{t}\right)$ dependence. In the
case of the stationary perturbations with $\gamma=0$ the integral
over $\widehat{t}$ is easily calculated and we obtain for times
$t\gg \tau_{correlation}\sim (\omega_{d}-lv_{0}'x)^{-1}$ the well
known result\cite{Monin}
\begin{eqnarray}
&\displaystyle K_{yy}\left(t\right)=\frac{c^{2}}{B^{2}}Re \int
dk_{\perp}\int
dl\left|\phi\left(k_{\perp},l,0\right)\right|^{2}\frac{C_{1}\left(k_{\perp},l,\widehat{t},
x\right)}
{\left(\omega_{d}\left(k_{\perp},l\right)-lv'_{0}x\right)^{2}}\nonumber
\\  &\displaystyle
\times\left(\frac{2}{3}\left(lv'_{0}\right)^{2}t^{3}-2k_{\perp}lv'_{0}t^{2}+2k^{2}_{\perp}t\right)
,\label{46}
\end{eqnarray}
which displays the effect of the anisotropic dispersion conditioned
by flow shear, observed in the laboratory frame of reference -
dispersion increases much faster along flow than in the direction of
the flow shear.

\section*{VI.Conclusions}
In this paper we present the results of the analytical
investigations of the linear and nonlinear evolution of drift modes
in shear flows with time dependent velocity shear. The
transformation (\ref{3}) to convective frame of reference gives an
opportunity to obtain the exact solution  for time dependent flow
shear. The investigations demonstrate that there are different time
scales in the dynamics of drift waves in shear flows. Right with
convective variables $\xi, \eta$ (\ref{3}) during times, for which
$\left|\alpha\left(t\right)v'_{0}l\right|\rho_{s}\ll 1$ (or, for the
stationary velocity shear, when $t\ll
\left(l\rho_{s}v'_{0}\right)^{-1}$) the solution (\ref{9}) of the
modal type is obtained. That modal solution already during early
times, for which $\left|\alpha\left(t\right)v'_{0}\right|> 1$, is
observed in the laboratory frame as a sheared mode with
time-dependent wave number and frequency. In this time domain in the
convective variables the dispersive properties of the plasma
displacements in drift turbulence are the same as in plasmas without
shear flow. Using the developed two-time scale procedure of the
calculation of the dispersion tensor of random displacements of the
plasma we obtain the renormalized form (\ref{38}) of the solution
for the electrostatic potential of the unstable drift waves. This
solution gives the nonlinear integral balance equation (\ref{39}),
which determines the level of drift turbulence in the steady state,
resulted from random turbulent motion of plasma in the unstable
drift turbulence. Balance equation (\ref{39}) governs the process
and level of the drift turbulence saturation. It does not include
any effects associated with flow shear over times during which the
solution (\ref{9}) is valid.

The effect of the anisotropic dispersion of plasma displacements is
presented by Eq.(\ref{46}). It presents the variance of the plasma
displacements resulted from the modes of the ordinary modal form
(\ref{9}), \textit{observed in the laboratory frame}. This effect
has nothing in common with effect of the "enhanced suppression of
the instability by shear flow". The saturation of the instability,
as it was demonstrated above, resulted from the balance (\ref{39})
of the growth rate and nonlinear damping in the frame of reference,
where the solution in the ordinary modal form is determined.

This result may be extended to other fluid models of plasma. The
fluid equations, obtained in drift approximation, in which all
nonlinearities other than $E\times B$ nonlinearity are ignored, do
not include among the nonlinear mechanisms of the instability
saturation the process of the enhanced nonlinear decorrelation by
velocity shear. It is important to note, that the same conclusions
concerning "suppression of the instability by the enhanced nonlinear
decorrelation by velocity shear" are completely applicable to
stratified shear flows of incompressible fluids, where the nonlinear
convective derivative
\begin{eqnarray}
\mathbf{v}\cdot\bigtriangledown=\frac{\partial\psi}{\partial
z}\frac{\partial}{\partial x}- \frac{\partial\psi}{\partial
x}\frac{\partial}{\partial z}\label{47}
\end{eqnarray}
presented in terms of the stream function
$\mathbf{\psi}=\mathbf{e}_{y}\psi$, with which
$\mathbf{v}=\bigtriangledown\times \mathbf{\psi}$ and where basic
flow along axis $x$ with velocity gradient along axis $z$ is
assumed. Eq.(\ref{47}) also conserves its form under transformation
(\ref{3}).

The process of the instabilities suppression by the enhanced
nonlinear decorrelation by velocity shear is treated appropriately
with nonlinear kinetic (gyrokinetic) theory of plasma with shear
flow (see, for example, nonlinear gyrokinetic turbulence simulations
of $E\times B$ shear quenching of transport in
Refs.\cite{Waltz1995}, \cite{Waltz1998}; we mention also
Ref.\cite{Kinsey}, \cite{Hauff, Mikhailenko-2009} as examples of a
recent works). We have obtained analytically  in
Ref.\cite{Mikhailenko-2009} that nonlinear effect of the suppression
instability by shear flow deduced straightforwardly in
Vlasov-Poisson system of equations for the magnetized plasma as the
effect of the enhanced by shear flow nonlinear resonance broadening
and it may be dominant in the processes of the instabilities
saturation.
\section*{Acknowledgements.}
This paper is the extended version of the report \cite{Mikhailenko}
P2-07 "Temporal evolution of drift-type modes in plasma flows with
strong time-dependent velocity shear" of the same authors, presented
to 4-th IAEA Technical Meeting on Theory of Plasma Instabilities
(Kyoto University, Japan, May 18-20, 2009). One of the authors (VSM)
would like to thank the International Atomic Energy Agency for
financial support of his visit to this Meeting.

\end{document}